\documentclass[a4paper,twocolumn,allowtoday,nopdfoutputerror]{quantumarticle}

\usepackage{amsmath}
\usepackage{amssymb}
\usepackage{graphicx}
\usepackage{float}
\usepackage{algorithm, algorithmic}
\usepackage{bm}
\usepackage{hyperref}

\begin{document}

\title{Correcting quantum errors one gradient step at a time}

\author{Manav Seksaria}
\affiliation{Center for Quantum Information, Communication and Computing, Indian Institute of Technology Madras, Chennai, India}
\email{manavs@ee.iitm.ac.in}

\author{Anil Prabhakar}
\affiliation{Department of Electrical Engineering, Indian Institute of Technology Madras, Chennai, India}
\email{anilpr@ee.iitm.ac.in}

\date{\today}

\begin{abstract}
Probabilistic approximate quantum error correction (PAQEC) lets a recovery fail with some probability, in exchange for higher fidelity when it succeeds.
Prior PAQEC code searches keep the codewords orthonormal with penalty terms and tune the code without its best recovery.
We propose a gradient method with three new parts.
First, the codewords move on the complex Stiefel manifold, so they are orthonormal at every step without penalties.
Second, every code is trained against its best recovery, which we solve for at each step rather than learn.
Third, a fast Petz recovery that we can differentiate even for symmetric codes gets the search close, and the exact semidefinite program finishes it.
We train a single five-qubit code over generalized amplitude damping noise.
At weak noise ($\gamma = 0.1$, $p = 0.05$) and equal success probability, it has $23$ times lower infidelity than the permutation-invariant code of prior work, and $9$ times lower than that work's numerically optimized code.
At equal fidelity it succeeds $91\%$ of the time, against $76\%$ for the permutation-invariant code.
\end{abstract}

\maketitle

\section{Introduction}
Machine learning has been applied across quantum error correction (QEC): to find codes~\cite{Jayashankar2023, Meyer2025, Olle2024}, to design encoders~\cite{Meyer2025, Olle2024} and to optimize recoveries~\cite{Biswas2024}. Most approaches assume some physical structure, to shrink the search space or to help convergence. Such assumptions restrict the set of codes the search can reach.

Probabilistic approximate quantum error correction (PAQEC)~\cite{Dutta2026} unifies approximate QEC (AQEC)~\cite{Jayashankar2023} and probabilistic QEC (PQEC) in one framework.
AQEC relaxes the Knill--Laflamme conditions~\cite{Jayashankar2023} and accepts small errors in exchange for more practical codes.
PQEC lets the recovery succeed only with some probability, but with high fidelity when it does.
PAQEC allows both at once: approximate error-correction conditions and a trace-non-increasing (probabilistic) recovery.
Within PAQEC, prior work~\cite{Dutta2026} constructs codes analytically or by a numerical search with orthonormality penalties, and finds the best recovery separately. Its code search does not use the best recovery while it tunes the code.

In this work we search for better PAQEC codes with minimal assumptions. The codewords are a point on the complex Stiefel manifold, so orthonormality holds by construction. For every code we solve for its best recovery, and take gradient steps on the code to increase the entanglement fidelity~\cite{Jayashankar2023} that this recovery reaches.

\section{Manifold Optimization}\label{sec:manifold}

The codewords must stay orthonormal throughout descent.
The code search of~\cite{Dutta2026} enforces this with penalty terms, whose weights must be tuned causing it's gradients compete with the fidelity gradient.
We avoid penalties altogether by optimizing directly on the set of orthonormal codeword frames, with $Q$ the number of physical qubits.

\subsection{Complex Stiefel Manifold}

Stack the $k$ codewords as columns of a matrix $X\in\mathbb{C}^{n\times k}$, where $n=2^Q$.
Orthonormality of the codewords is the single constraint $X^\dagger X = I_k$, and the set of all such matrices is the \emph{complex Stiefel manifold}
\begin{align}
    \mathrm{St}(n,k;\mathbb{C}) = \bigl\{ X\in\mathbb{C}^{n\times k} : X^\dagger X = I_k \bigr\},
\end{align}
a smooth real manifold of dimension $2nk - k^2$ embedded in $\mathbb{C}^{n\times k}$.
For five qubits and two codewords, $(n, k) = (32, 2)$ and the manifold has dimension $124$. Every point of it is an orthonormal tuple, so no penalty term is needed.

\subsection{Riemannian Descent}
We use the standard Riemannian machinery for the Stiefel manifold~\cite{Absil2008}. The tangent space at $X$ is
\begin{align}
    T_X\mathrm{St} = \bigl\{ \Delta\in\mathbb{C}^{n\times k} : X^\dagger\Delta + \Delta^\dagger X = 0 \bigr\}.
\end{align}
The Riemannian gradient is the projection onto the tangent space of the Euclidean gradient $G$, which automatic differentiation computes,
\begin{multline}\label{eq:riemannian_grad}
    \mathrm{proj}_X(G) = G - X\,\mathrm{sym}_H(X^\dagger G), \\
    \mathrm{sym}_H(A) = \frac{A + A^\dagger}{2}.
\end{multline}
A step $U$ in the tangent space moves $X$ to $X+U$, which is off the manifold. A QR decomposition retracts it back onto $\mathrm{St}(n,k;\mathbb{C})$~\cite{Absil2008},
\begin{align}
    \mathrm{retr}_X(U):\quad X + U = QR \implies \mathrm{retr}_X(U) = Q.
\end{align}
We use Riemannian Adam~\cite{Becigneul2019}, which applies the Adam moment accumulation to the Riemannian gradient and then retracts, so each parameter keeps an adaptive learning rate and the iterate stays on the manifold.

\section{Optimization}\label{sec:opt}

We optimize five-qubit codes under generalized amplitude damping (GAD)~\cite{Dutta2026}, which damps each qubit at rate $\gamma$ and excites it with relative weight $p$. We train over a grid of noise points and, for every code, solve for its best recovery.

\subsection{Fidelity and Success Probability}\label{sec:opt_defs}

We use the post-selected entanglement fidelity and success probability of~\cite{Dutta2026}. The physical space $\mathcal{H}_C$ of $Q$ qubits has dimension $2^Q$, the logical space $\mathcal{H}_S$ has dimension $k$, and the noise channel $\mathcal{E}$ has Kraus operators $\{E_i\}$. A recovery is a set of operators $\{R_j\}$ from $\mathcal{H}_C$ to $\mathcal{H}_S$ with $\sum_j R_j^\dagger R_j \preceq I$, the deficit being the failure outcome. The success probability $\bar p$ is the trace of the recovered state, and $F_{\text{ent}}$ is the entanglement fidelity~\cite{Jayashankar2023} of the recovered state renormalized by $\bar p$.

The fidelity is invariant under $R_j \mapsto sR_j$ for any $s>0$, so keeping fewer outcomes can raise it. We therefore report $\bar p$ with every fidelity and compare codes at equal $\bar p$. Our reference is the $[5,1]$ permutation-invariant (PI) code of~\cite{Dutta2026}, a member of the wider class of permutation-invariant codes~\cite{Pollatsek2004, OuyangChao2020}.

\subsection{Noise Grid}\label{sec:opt_grid}

A code trained at one noise point is fitted to it, so we average the loss $1 - F_{\text{ent}}$ over a batch drawn from the grid $\gamma, p \in \{0, 0.05, \ldots, 0.5\}$, $121$ points in all for our test case. The Kraus operators of each grid point are built once and reused. We assume the noise is independent and identical on every qubit (IID), so the channel is a tensor product of $Q$ single-qubit GAD channels with four Kraus operators each. Applying these one qubit at a time, rather than expanding them into $4^Q$ operators, replaces $1024$ dense $32\times32$ products by $20$ small ones at five qubits, which makes a batch of grid points cheap.

Because the fidelity is post-selected, a recovery can raise it by succeeding less often (Section~\ref{sec:opt_defs}). The solved recovery fixes the success probability directly through $c$, but the Petz recovery does not, so training against it could buy fidelity by lowering $\bar p$. A floor penalty $\mu\max(0,\,c_0-\bar p)^2$ with weight $\mu$ blocks this by holding $\bar p$ above $c_0$. We take $c_0 = \bar p_{\text{PI}}$, the success probability of the $[5,1]$ PI code, so that every comparison is at equal $\bar p$. This forces us to perform \textit{at least} as well as the PI code.

\subsection{Solved Recovery}\label{sec:opt_envelope}

For each code $X$ we want the recovery that gives the highest fidelity. We call it $R^\ast(X)$, and the fidelity it reaches $F^{\max}(X)$,
\begin{align}\label{eq:Fstar}
    F^{\max}(X) = \max_{R\in\mathcal{R}} F_{\text{ent}}(X, R),
\end{align}
where $\mathcal{R}$ is the set of all physically allowed recoveries (completely positive and trace non-increasing). We improve the code by gradient steps on $F^{\max}$. This looks hard, because $R^\ast$ changes whenever $X$ changes. Danskin's theorem~\cite{Danskin1967} removes the problem. If $R^\ast$ is unique, then
\begin{align}\label{eq:danskin}
    \nabla_X F^{\max}(X) = \nabla_X F_{\text{ent}}(X, R)\big|_{R = R^\ast(X)},
\end{align}
so we can hold the recovery fixed at its best and differentiate the code alone. The reason is that $R^\ast$ is already the best recovery, so changing it slightly does not change the fidelity to first order. Only the change in the code counts. This holds for one small step, and we solve for $R^\ast$ again after every step, so the error does not build up. The recovery is therefore never a parameter we train. A see-saw~\cite{Fletcher2007thesis}, which alternates between solving for the recovery and for the code, has the same stationary points.

We use two recoveries, for speed. Training takes many steps, and each step visits many noise points. The solved recovery is the best possible, but each solve is an SDP that takes seconds, so training on it alone would take hours. The Petz recovery is a closed formula and takes a fraction of a second. It is also near optimal: its infidelity is at most about twice that of the best recovery~\cite{Barnum2002, Ng2010}, so a code that is good under Petz is close to good under the solved recovery. We therefore train against Petz to get close quickly, then switch to the solved recovery to finish.

The first is the Petz recovery in its transpose-channel form~\cite{Jayashankar2023}. It needs a set of errors to undo. The order of an error is the number of qubits it damps or excites. We keep only the errors up to a chosen truncation order, which keeps the recovery small. Let $S$ be this set and $P = XX^\dagger$ the projector onto the code space. Eq.~\eqref{eq:petz} builds the recovery,
\begin{align}\label{eq:petz}
    \mathcal{E}(P) = \sum_{i\in S} E_i P E_i^\dagger,
    \quad
    R_i = X^\dagger E_i^\dagger\, \mathcal{E}(P)^{-1/2},
\end{align}
where $\mathcal{E}$ keeps only the errors in $S$ and $R_i$ undoes the error $E_i$. Petz is not the best recovery, so Eq.~\eqref{eq:danskin} does not apply to it. We therefore differentiate through $R_i$ rather than freezing it. The derivative of $\mathcal{E}(P)^{-1/2}$ divides by the gaps between its eigenvalues~\cite{Higham2008}. Symmetric codes have equal eigenvalues, where this gives $0/0$. There we use the slope of $\lambda^{-1/2}$ instead, which is the limit as the two eigenvalues meet.

The second is the optimal recovery~\cite{Fletcher2007thesis}, which we call the solved recovery. It is the recovery that Eq.~\eqref{eq:Fstar} asks for. We find it with the semidefinite program (SDP) of~\cite{Dutta2026}, whose variable is the Choi matrix $J$ of the recovery. The fidelity is post-selected, so we fix the success probability at a target $c$ and ask for the best fidelity at that $c$. Two matrices built from the code enter, $C(X)$ and $B(X)$, with $\mathrm{Tr}[JC(X)] = \bar p\,F_{\text{ent}}$ and $\mathrm{Tr}[JB(X)] = \bar p$. Both are linear in $J$. Following Eq.~\eqref{eq:frontier} specify the SDP and the fidelity it reaches,
\begin{multline}\label{eq:frontier}
    N(c) = \max_J \bigl\{\,\mathrm{Tr}[JC(X)] \;\big|\; J\succeq0,\ \mathrm{Tr}_{\mathcal{H}_S}J \preceq I, \\
    \mathrm{Tr}[JB(X)] = c \,\bigr\},
    \qquad
    F^\ast(c) = \frac{N(c)}{c}.
\end{multline}
The curve $F^\ast(c)$ is non-increasing in $c$~\cite{Dutta2026}. The gradient at fixed $c$ again follows Danskin's theorem, now applied to the Lagrangian of the program. The constraint $\mathrm{Tr}[JB(X)] = c$ moves with $X$, and its dual variable $\nu^\ast = \mathrm{d}N/\mathrm{d}c$ is the cost of success probability. With $J^\ast$ as the maximizer, Eq.~\eqref{eq:danskin_c} gives the gradient,
\begin{align}\label{eq:danskin_c}
    \nabla_X F^\ast(c) = \frac{1}{c}\,\nabla_X\Bigl(\mathrm{Tr}[J^\ast C(X)] - \nu^\ast\,\mathrm{Tr}[J^\ast B(X)]\Bigr).
\end{align}
For a five-qubit code the recovery maps $32$ dimensions to $2$, so $J$ is $64\times64$. Algorithm~\ref{alg:envelope} lists the training loop of $T$ steps with learning rate $\eta$. Each step draws a batch of grid points, solves one recovery per point, averages the gradients, and then projects and retracts as in Section~\ref{sec:manifold}.

\begin{algorithm}[H]
\caption{Descent Against the Solved Recovery}\label{alg:envelope}
\begin{algorithmic}[1]
\STATE Init $X_0 \in \mathrm{St}(n,k;\mathbb{C})$
\FOR{$t = 0, 1, \ldots, T-1$}
    \STATE Draw a batch $\mathcal{B}$ of points from the grid
    \STATE \textit{}
    \FOR{each $(\gamma, p) \in \mathcal{B}$}
        \STATE Solve Eq.~\eqref{eq:frontier} for $(J^\ast, \nu^\ast)$
        \STATE $G \leftarrow -\nabla_X F^\ast(c)$ from Eq.~\eqref{eq:danskin_c}
    \ENDFOR
    \STATE \textit{}
    \STATE $\tilde{G} \leftarrow \mathrm{proj}_{X_t}$
    \STATE $X_{t+1} \leftarrow \mathrm{retr}_{X_t}\!\bigl(-\eta\,\tilde{G}\bigr)$
\ENDFOR
\end{algorithmic}
\end{algorithm}

\subsection{Results}\label{sec:opt_results}

We optimize under GAD over the grid of Section~\ref{sec:opt_grid} in two stages.
\begin{itemize}
    \item \textbf{Stage one} runs $750$ steps against the Petz recovery of Eq.~\eqref{eq:petz} from a random $X_0$, with learning rate $\eta = 0.02$ decayed to $0.001$ after step $500$.
    \item \textbf{Stage two} runs $150$ steps of Algorithm~\ref{alg:envelope} against the solved recovery at $c = \bar p_{\text{PI}}$, with $\eta = 0.005$ decayed to $0.001$ after step $100$.
\end{itemize}
One solve of Eq.~\eqref{eq:frontier} takes about three seconds with CVXPY~\cite{Diamond2016} and SCS~\cite{ODonoghue2016}.

Ignoring $\bar p$ is misleading. At $\gamma=0.1$, $p=0.05$, a random code with its best recovery reaches $F_{\text{ent}} = 0.9992$, above the PI code's $0.9991$. But it succeeds only $18\%$ of the time, against $67\%$ for the PI code. We therefore always compare at equal $\bar p$.

\begin{figure}[t]
\includegraphics[width=\columnwidth]{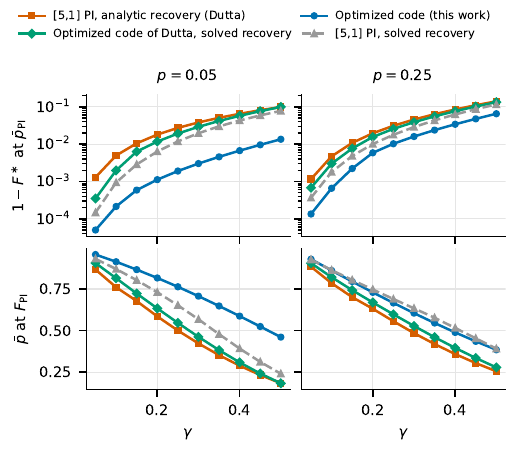}
\caption{Our optimized code beats the PI code of~\cite{Dutta2026} at every noise point. Top, infidelity at the PI code's success probability $\bar p_{\text{PI}}$. Bottom, success probability at the PI code's fidelity $F_{\text{PI}}$. Orange, the PI code with its analytic recovery, which sets $\bar p_{\text{PI}}$ and $F_{\text{PI}}$. Green, the numerically optimized code of~\cite{Dutta2026} with its solved recovery. Blue, our optimized code with its solved recovery. Dashed grey, the PI code with its solved recovery.}\label{fig:curves}
\end{figure}

We evaluate every code with its own solved recovery, so the comparison is between codes and not between recoveries. The analytic recovery of the $[5,1]$ PI code is the recovery that~\cite{Dutta2026} constructs for it. With $F_{\text{PI}}$ the fidelity of the PI code with this recovery, we report two quantities at each grid point.
\begin{itemize}
    \item The infidelity $1 - F^\ast$ at $\bar p_{\text{PI}}$.
    \item The success probability attainable at $F_{\text{PI}}$.
\end{itemize}
Figure~\ref{fig:curves} shows both against $\gamma$ at $p = 0.05$ and $p = 0.25$. Compare the blue line with the orange line. At every point the optimized code outperforms the PI code with its analytic recovery in both fidelity and success rate. The green line is the numerically optimized code of~\cite{Dutta2026}, given our solved recovery. The dashed grey line gives the PI code our solved recovery as well for sake of completeness.

\section{Conclusion}
We have proposed a gradient method for finding PAQEC codes, with three new parts. The codewords move on the complex Stiefel manifold, so they stay orthonormal without penalty terms. Every code is trained against its best recovery, and Danskin's theorem gives the exact gradient with that recovery held fixed, so the recovery is solved for and never learned. A Petz recovery that stays differentiable for symmetric codes makes training fast, and the SDP finishes it.

Trained once over the $121$-point GAD grid, our five-qubit code beats both the PI code and the numerically optimized code of~\cite{Dutta2026} at every point we tested. At weak noise ($\gamma = 0.1$, $p = 0.05$) and equal success probability, its infidelity is $23$ times lower than the PI code's and $9$ times lower than the optimized code's. At equal fidelity it succeeds $91\%$ of the time, against $76\%$ for the PI code. Even when the PI code also gets the best recovery, our code has $4.4$ times lower infidelity there, so the gain comes from the code and not only from the recovery.

Several directions remain open. Larger codes and more logical qubits need cheaper recoveries, since the SDP grows quickly with the number of qubits. Other noise models, such as noise measured on hardware, only change the channel and not the method. Correlated noise would lose the per-qubit speedup of Section~\ref{sec:opt_grid}. Finally, our codes are trained and tested on the same noise grid, so testing them on held-out noise points would show how well they generalize.

\section{Acknowledgements}
M.S. thanks Rishabh Patra for teaching him machine learning. This work received financial support from the Mphasis F1 Foundation, the National Quantum Mission initiative of the Department of Science \& Technology.

\bibliographystyle{quantum}
\bibliography{ref}

\appendix

\section{Optimized Codewords}\label{sec:codes}

The optimized codewords below are rounded to four decimal places.
Each is a 32-dimensional complex vector ($2^5$ amplitudes) arranged in the computational basis order $|00000\rangle,\ldots,|11111\rangle$, written as a row vector.

\onecolumn
For the GAD grid of Section~\ref{sec:opt_grid}, trained over $\gamma, p \in \{0, 0.05, \ldots, 0.5\}$,

\begin{equation}
\scriptsize
\begin{aligned}
|0_L\rangle = [&-0.0757 - 0.2317i, -0.1160 + 0.0063i, 0.0509 + 0.1032i, 0.1054 - 0.0013i, -0.0536 + 0.1194i, -0.1854 + 0.2010i,\\
&-0.2324 + 0.1330i, 0.0809 + 0.0306i, -0.0138 + 0.1306i, -0.1151 - 0.2425i, 0.2734 + 0.0075i, 0.0868 + 0.0008i,\\
&-0.1534 - 0.2273i, -0.0719 - 0.0383i, -0.0058 + 0.0828i, 0.1100 + 0.0380i, 0.2455 - 0.1050i, 0.0344 - 0.2538i,\\
&-0.2507 + 0.0661i, 0.1691 - 0.2239i, -0.1018 + 0.0025i, 0.1718 - 0.0188i, -0.0683 + 0.1765i, 0.0402 - 0.0731i,\\
&-0.0967 + 0.0324i, -0.1715 - 0.0806i, -0.1289 - 0.1154i, 0.0138 - 0.0827i, 0.0581 - 0.0027i, -0.0157 + 0.0386i,\\
&0.0415 - 0.0006i, -0.0199 + 0.0373i]\\
|1_L\rangle = [&0.1007 - 0.1880i, -0.0686 - 0.0759i, -0.0426 + 0.0920i, 0.0448 + 0.0673i, -0.1483 + 0.2146i, -0.2056 - 0.0345i,\\
&-0.1984 - 0.0861i, 0.0202 + 0.0670i, -0.0741 + 0.2502i, 0.1002 - 0.1884i, 0.1123 + 0.1786i, 0.0422 + 0.0562i,\\
&-0.0363 + 0.2657i, -0.0492 - 0.0942i, -0.0663 + 0.0784i, 0.0268 + 0.0860i, -0.2926 - 0.1625i, -0.1464 + 0.1529i,\\
&0.1959 + 0.0724i, 0.2330 - 0.0072i, 0.0767 + 0.0919i, -0.1650 - 0.1609i, 0.2108 - 0.0462i, -0.0487 + 0.0298i,\\
&0.1020 + 0.0645i, 0.0304 + 0.2127i, -0.0105 + 0.2310i, -0.0364 + 0.0436i, -0.0458 - 0.0472i, 0.0398 - 0.0168i,\\
&-0.0309 - 0.0300i, 0.0322 - 0.0191i].
\end{aligned}
\end{equation}

\end{document}